\documentclass[a4paper,journal,10pt,twocolumn]{IEEEtran}
\usepackage{acronym}
\usepackage{xcolor}
\usepackage{graphicx}
\usepackage{url}
\usepackage{balance}

\usepackage[colorinlistoftodos,prependcaption,textsize=small]{todonotes}

\acrodef{AI}{Artificial Intelligence}
\acrodef{RF}{Radio Frequency}
\acrodef{RFF}{Radio Frequency Fingerprinting}
\acrodef{SIGINT}{Signal Intelligence}
\acrodef{SDR}{Software-Defined Radio}
\acrodef{IQ}{In-Phase Quadrature}
\acrodef{ML}{Machine Learning}
\acrodef{DL}{Deep Learning}
\acrodef{SpecInt}{Cyber Spectrum Intelligence}
\acrodef{SNR}{Signal to Noise Ratio}
\acrodef{WSI}{Wireless Signal Identification}
\acrodef{AMR}{Automatic Modulation Recognition}
\acrodef{RSS}{Received Signal Strength}
\acrodef{ToA}{Time of Arrival}
\acrodef{CI}{Critical Infrastructure}
\acrodef{SNR}{Signal-to-Noise Ratio}
\acrodef{COTS}{Commercial Off-The-Shelf Devices}
\acrodef{BER}{Bit-Error Rate}

\newcommand{\specint}{{\em SpecInt}}

\title{Cyber Spectrum Intelligence: Security Applications, Challenges and Road Ahead}

\author{
    \IEEEauthorblockN{Savio Sciancalepore\IEEEauthorrefmark{1}, Gabriele Oligeri\IEEEauthorrefmark{2}}\\
    \IEEEauthorblockA{\IEEEauthorrefmark{1}Eindhoven University of Technology, Eindhoven, Netherlands.\\ s.sciancalepore@tue.nl}\\
    \IEEEauthorblockA{\IEEEauthorrefmark{2}College of Science and Engineering, Hamad Bin Khalifa University, Doha, Qatar.\\ goligeri@hbku.edu.qa}
}

\begin{document}

\maketitle

\begin{abstract}
Cyber Spectrum Intelligence (SpecInt) is emerging as a concept that extends beyond basic {\em spectrum sensing} and {\em signal intelligence} to encompass a broader set of capabilities and technologies aimed at monitoring the use of the radio spectrum and extracting information. SpecInt merges traditional spectrum sensing techniques with Artificial Intelligence (AI) and parallel processing to enhance the ability to extract and correlate simultaneous events occurring on various frequencies, allowing for a new wave of intelligence applications.

This paper provides an overview of the emerging SpecInt research area, characterizing the system architecture and the most relevant applications for cyber-physical security. We identify five subcategories of spectrum intelligence for cyber-physical security, encompassing Device Intelligence, Channel Intelligence, Location Intelligence, Communication Intelligence, and Ambient Intelligence. We also provide preliminary results based on an experimental testbed showing the viability, feasibility, and potential of this emerging application area. Finally, we point out current research challenges and future directions paving the way for further research in this domain.

\end{abstract}

\section{Introduction}
\label{sec:introduction}

Recent technological advancements in \acp{SDR}, \ac{AI}, and (parallel) processing capabilities are enhancing the current state of the art in several communication-related research domains, enabling new solutions in the areas of \emph{spectrum sensing}, \emph{signal intelligence}, and \emph{cognitive radio networking}, to name a few~\cite{paul2016_access}. The commercial market also shows an increasing interest in such technologies, as highlighted by the expected increase of 15.3\% in the market size only for cognitive radio applications by 2030~\cite{gvr_market}.

In this context, many scientific contributions have demonstrated effective capabilities of inferring valuable information using data collected from the \ac{RF} spectrum. For example, research in \ac{RFF} demonstrated the viability of identifying specific devices by analyzing their wireless signals through \ac{AI} solutions~\cite{alhazbi2024_iwcmc}, while other scientists have obtained remarkable results in detecting communications and classifying their main characteristics (e.g. modulation) through advanced analysis of the \ac{RF} spectrum~\cite{zhang2022_dsp}. Moreover, \ac{RF} spectrum analysis is more and more considered for anomaly detection in specific wireless scenarios, e.g., satellites, where distinguishing propagation phenomena can ease the localization of the transmitting devices~\cite{paul2016_access, oligeri2024_sac}.

\begin{figure}[t]
  \centering
  \includegraphics[width=\columnwidth, angle = 0,trim = 25mm 0mm 25mm 0mm]{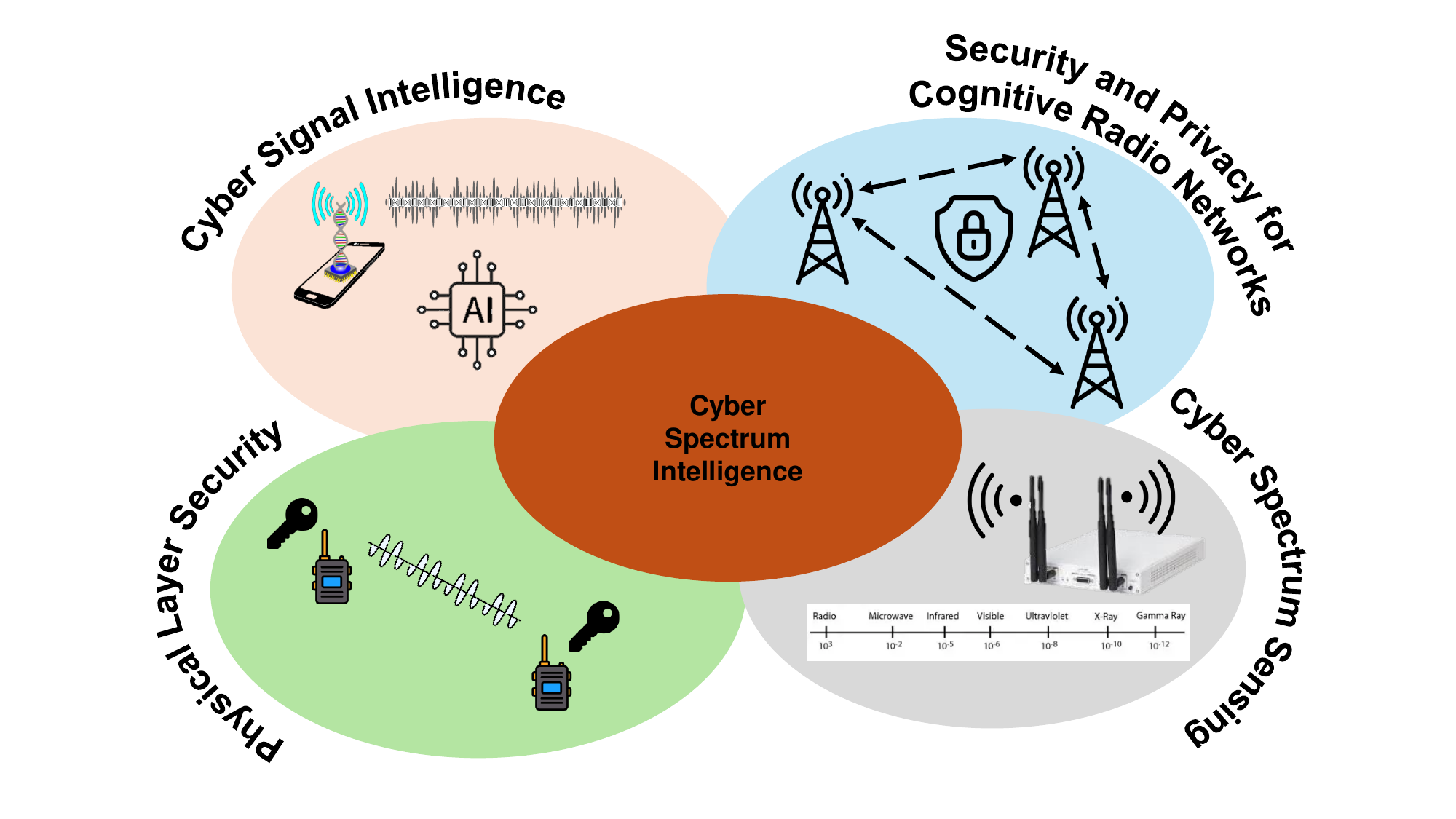}
  \caption{The {\em Cyber-Spectrum Intelligence} paradigm emerges at the intersection between cyber-signal intelligence, security and privacy applications of cognitive radio networks, cyber spectrum sensing and physical-layer security.} 
  \label{fig:general}
\end{figure}

However, the research in this area mostly aims to improve communication quality. For example, recent research in signal intelligence aims at adapting the modulation chain at the receiver to enable timely communication~\cite{zhang2022_dsp}; recent solutions for cognitive radio networks aim to improve channel allocation between primary and secondary users~\cite{zhang2024_css,dipietro2013_net}; solutions in the area of cyber spectrum sensing allow detecting humans moving across a radio link, thus allowing switching to less noisy channels~\cite{paul2016_access}; and finally, physical-layer techniques have been discussed in~\cite{huang2020_wcom} to improve the performance of 5G wireless communications. 
Radio spectrum sensing, cognitive radio networks, cyber-signal intelligence, and physical-layer security for security and privacy applications are all topics leveraging physical-layer information from the radio spectrum.
Although the performance of the solution mentioned above for enhancing communication quality is promising, we also see the potential to leverage physical-layer information from the radio spectrum for security and privacy purposes, i.e., {\em detecting any cyber and physical security event of interest in the close neighborhood of the transmitter-receiver link}.
Contributions on this topic are mostly scattered across various scientific communities, thus missing a unified view, system architecture, and overall systematization easing concrete advancements and commercialization. 
Overall, as shown in Fig.~\ref{fig:general}, we do see the potential for the development of a trans-disciplinary research domain, namely \acl{SpecInt} (\specint), aimed at the design and development of integrated techniques using advanced signal processing, \ac{AI}, and parallel processing to provide real-time cyber-physical security and privacy services.

The keyword \specint\ have emerged in recent years in both the scientific community and the market. Zhang et al. in~\cite{zhang2024_css} introduced the concept of Collaborative Spectrum Intelligence to aid and optimize the performance of devices operating in cognitive networks, helping them to identify available spectrum resources more efficiently. At the same time, Cisco commercialized access points supporting the feature \emph{Spectrum Intelligence}, enabling detection of common non-WiFi sources of interference for WiFi networks, e.g., baby monitors, microwaves, and Bluetooth networks~\cite{specint_cisco}.
However, to the best of our knowledge, there are no comprehensive scientific contributions and not even products systematizing the usage of physical-layer information of radio communications to provide: (i) enhanced security features, and (ii) detect security-sensitive events occurring in the surrounding environment.

{\bf Contribution.} In this paper, we introduce and systematize the paradigm of \emph{Cyber Spectrum Intelligence} (\specint) for cyber-physical security applications. We describe the system model, the entities involved, the constraints, and we identify five major cyber-physical security application domains where the deployment of \specint\ technologies could enhance the effectiveness of regular operations: device intelligence, channel intelligence, location intelligence, communication intelligence and ambient intelligence. We also provide preliminary results based on real-world experiments showing the feasibility of the most relevant identified applications. Finally, we unveil current challenges toward the deployment of reliable and effective \specint\ applications, laying down a future research agenda in this domain. To the best of our knowledge, this paper is the first to define the concept of \specint\ for cyber-physical security and to systematize knowledge in the domain.

{\bf Roadmap.} This paper is organized as follows. Sec.~\ref{sec:architecture} introduces the system architecture of \specint, Sec.~\ref{sec:applications} provides an overview of the main applications of \specint\ for cyber-physical security, as well as preliminary results supporting our vision, Sec.~\ref{sec:future} points out research challenges and future research directions, and finally, Sec.~\ref{sec:concl} concludes the paper.

\section{General System Architecture}
\label{sec:architecture}
\begin{figure}
  \centering
  \includegraphics[width=\columnwidth]{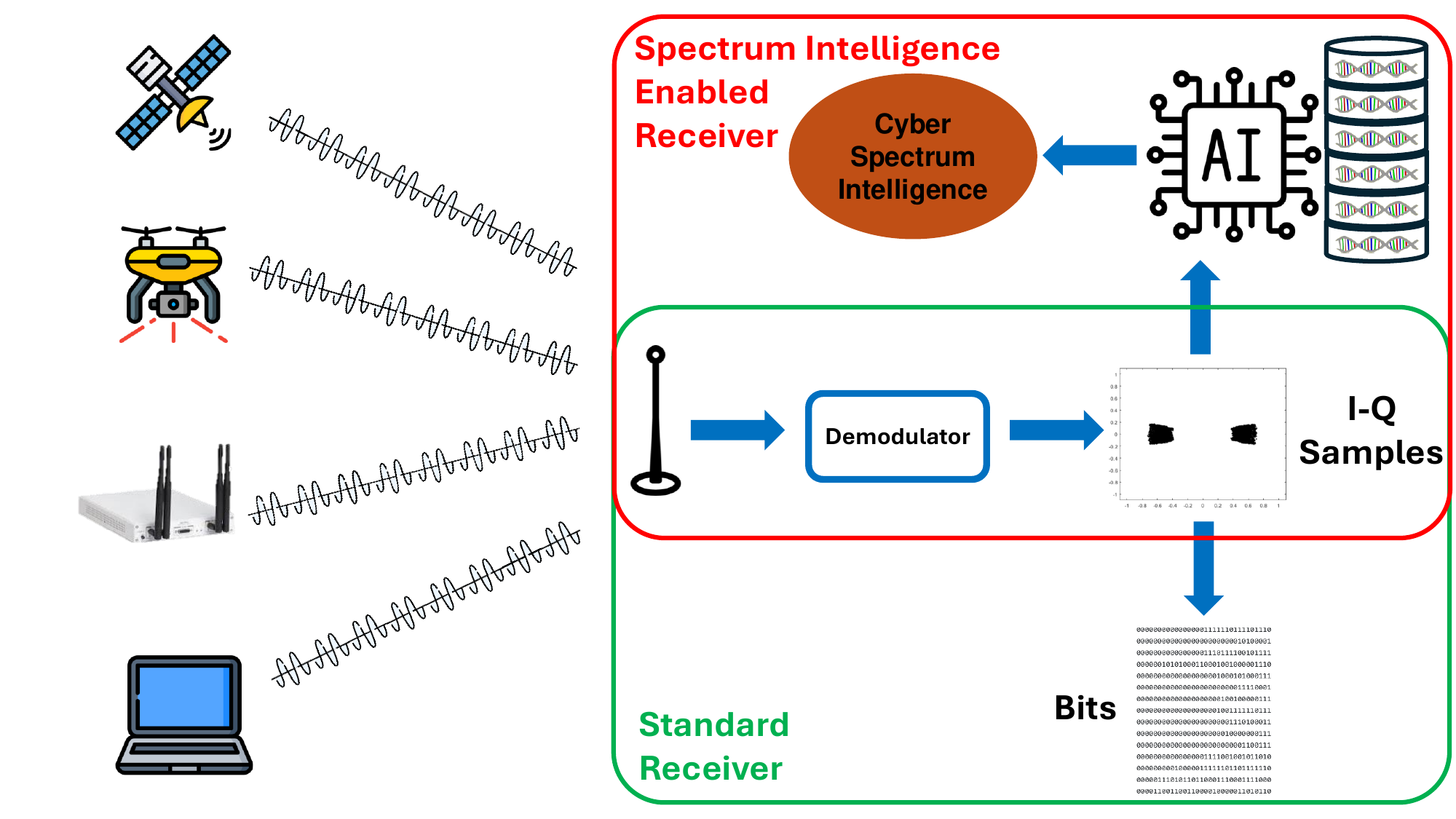}
  \caption{System Architecture enabling \specint. \ac{RF} transmitters send data through the radio spectrum. A (standard) receiver can collect such data and exploit the physical-layer information to enable \specint\ functionalities.}
  \label{fig:architecture}
\end{figure}

Figure~\ref{fig:architecture} shows the system architecture envisioned for \specint applications.

\specint applications require the deployment of \ac{RF} transceivers in the area of interest. While transmitters are simply required to access the radio spectrum and broadcast data, receivers undergo different requirements as a function of their \specint\ functionalities, i.e., red and green boxes in Fig.~\ref{fig:architecture}. A standard receiver only implements demodulation and data recovery from the radio spectrum, while a \specint\ receiver collects and processes physical-layer information from the radio spectrum. It is worth noting that a receiver can perform both operations, i.e., data recovery and \specint. Moreover, while the standard receiver should have a-priori knowledge about physical and logical parameters of the communication, e.g., carrier frequency, modulation, protocols, and crypto parameters, the \specint\ receiver may not be aware of the non-physical layer information (protocols and crypto secrets), and it can potentially recover any other information from the physical layer, e.g., communication bandwidth and modulation type. Indeed, the \specint\ receiver might not be part of the communication network. However, it can still perform \specint\ activities by exploiting the available information from the physical layer of the radio spectrum. 
Therefore, to be \specint\ compliant, a receiver should feature the ability to collect (and store) information from the physical layer of the radio spectrum and process it with a \ac{DL} algorithm. A common setup to implement a \specint-enabled receiver involves a \ac{SDR} and a host device to control the \ac{SDR}, e.g., running \emph{GNU Radio}, collecting information from the physical layer of the radio spectrum, and finally, processing them resorting to \ac{DL} techniques.

The \specint\ receiver should feature significant computational and processing capabilities, allowing the execution of state-of-the-art \ac{AI} algorithms for extracting information from the received data. Also, the \specint\ receiver should feature large storage space, which is necessary to host the large amount of data collected from the \ac{SDR}. Finally, we remark that \specint\ applications require at least one transmitter and one receiver. However, a notable difference compared to the system architecture of other physical-layer research domains (recall Fig.~\ref{fig:general}) is that the receiving system is not interested in recovering the information delivered by the transmitter, i.e., the bits which indeed might be encrypted. Instead, by analyzing the raw \ac{IQ} samples corresponding to such bits, it is possible to infer additional information, such as the identity of the device, the status of the channel, the location of the transmitter, specific features of the communication and, finally, events happening in the surrounding of the communication link. We discuss such applications in the following sections.

\section{Spectrum Intelligence Applications}
\label{sec:applications}

\subsection{Device intelligence}
\label{sec:devInt}
\specint\ applications falling into the area of \emph{Device Intelligence} allow to obtain insights about specific devices active in the area under monitoring. 

One notable research domain falling in this area is \ac{RFF}.
\ac{RFF}techniques are used to uniquely identify wireless devices exploiting the hardware differences of the transmitters, which create a distinctive ``fingerprint'' in the over-the-air signal~\cite{alhazbi2024_iwcmc}. \ac{RFF} is rooted in the idea that there are no two identical devices (at the electronic level), and such differences are eventually reflected in the over-the-air signal, enabling the receiver to distinguish the source of the signal (in a pool of candidates). These solutions are particularly relevant in the context of \ac{SIGINT}, where identifying, tracking, and counting signal sources is critical for surveillance and military operations. In fact, the analysis of the features extracted from the physical layer of the transmitted signals enables device identification and authentication even before the signal is decoded. This provides an additional layer of security in scenarios where traditional cryptographic methods cannot be deployed, because either they are too computationally intense or key distribution is problematic due to the considered scenario. \ac{RFF} can also be used for tracking devices in the wild when the receiver exploits the transmitter's fingerprint to infer its presence in a close neighborhood. Typical application scenario involves both security (tracking of devices such as drones) and privacy (tracking of smartphones to infer users' behavior). The former can involve malicious use of drones, e.g., unauthorized surveillance, smuggling, or even armed attacks, thus necessitating the development of methods for tracking and identifying these devices in real-time. \ac{RFF} capabilities, although useful for security applications, can be exploited also for tracking users without consent. Indeed, \ac{RFF} allows passive monitoring of devices’ signal, enabling to trace a user’s location or movements over time, even if the device hides the identifying information, e.g., MAC or IP address.

\begin{figure}
    \centering
    \includegraphics[width=.9\columnwidth, angle = 0,trim = 30mm 90mm 30mm 90mm]{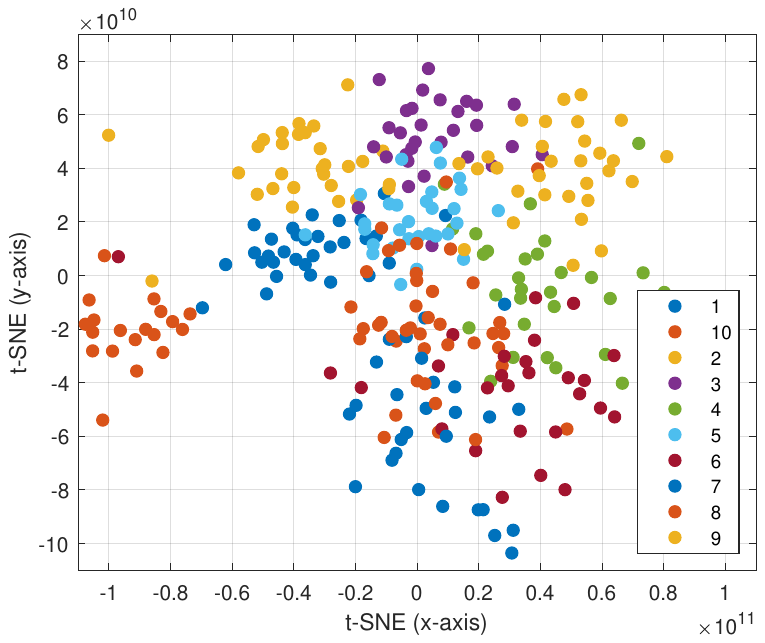}
    \caption{t-Distributed Stochastic Neighbor Embedding (t-SNE) analysis of the last layer of a {\em ResNet18} network considered for testing on a dataset of 10 transmitters. The accuracy of the classifier is close to 1.} 
    \label{fig:rff}
\end{figure}

A typical example proving the viability of device intelligence is the ability to discriminate devices in the radio spectrum by just resorting to physical-layer information. We collected data through a proof-of-concept deployment of a device intelligence application, using one \ac{SDR} USRP X310 as a receiver and ten different wireless \acp{SDR} transmitters USRP B210, all emitting the same wireless message. Then, we trained and tested a model to identify the specific transmitter based on the \ac{RF} transmissions of the devices, using the image-based RFF approach used in~\cite{alhazbi2023_acsac}. Figure~\ref{fig:rff} shows the t-Distributed Stochastic Neighbor Embedding (t-SNE) analysis of the testing data considering the mentioned classifier, which shows that the \ac{RF} profiles of the various transmitters are clearly recognizable and rarely overlapping, thus indicating the viability of performing \ac{RF} device intelligence.

\subsection{Channel intelligence} 
\label{sec:chanInt}
{\em Channel intelligence} refers to the systematic collection, analysis, and interpretation of data related to target frequency channels within the electromagnetic spectrum with the objective of identifying anomalies possibly indicating ongoing attacks. It involves understanding the usage patterns, signal characteristics, and occupancy of these channels to identify the presence of anomalies. In the context of cyber-physical security applications for \specint, channel intelligence is instrumental in safeguarding wireless communications against a range of radio spectrum threats. Since wireless technologies have become increasingly enabling for many critical applications, threats associated with malicious activities directly affecting the radio spectrum are becoming increasingly significant. Jamming is the most significant threat when considering denial of service at the physical layer of the radio spectrum, as it can degrade or completely block wireless communications, affecting everything from military operations to civilian telecommunications. Jamming is a malicious activity of transmitting radio signals (noise or other forms of interference) that disrupt or interfere with the legitimate communication of other devices by colliding (on the same communication channel) with the original signals. 

\begin{figure}
    \centering
    \includegraphics[width=.9\columnwidth, angle = 0,trim = 50mm 0mm 50mm 0mm]{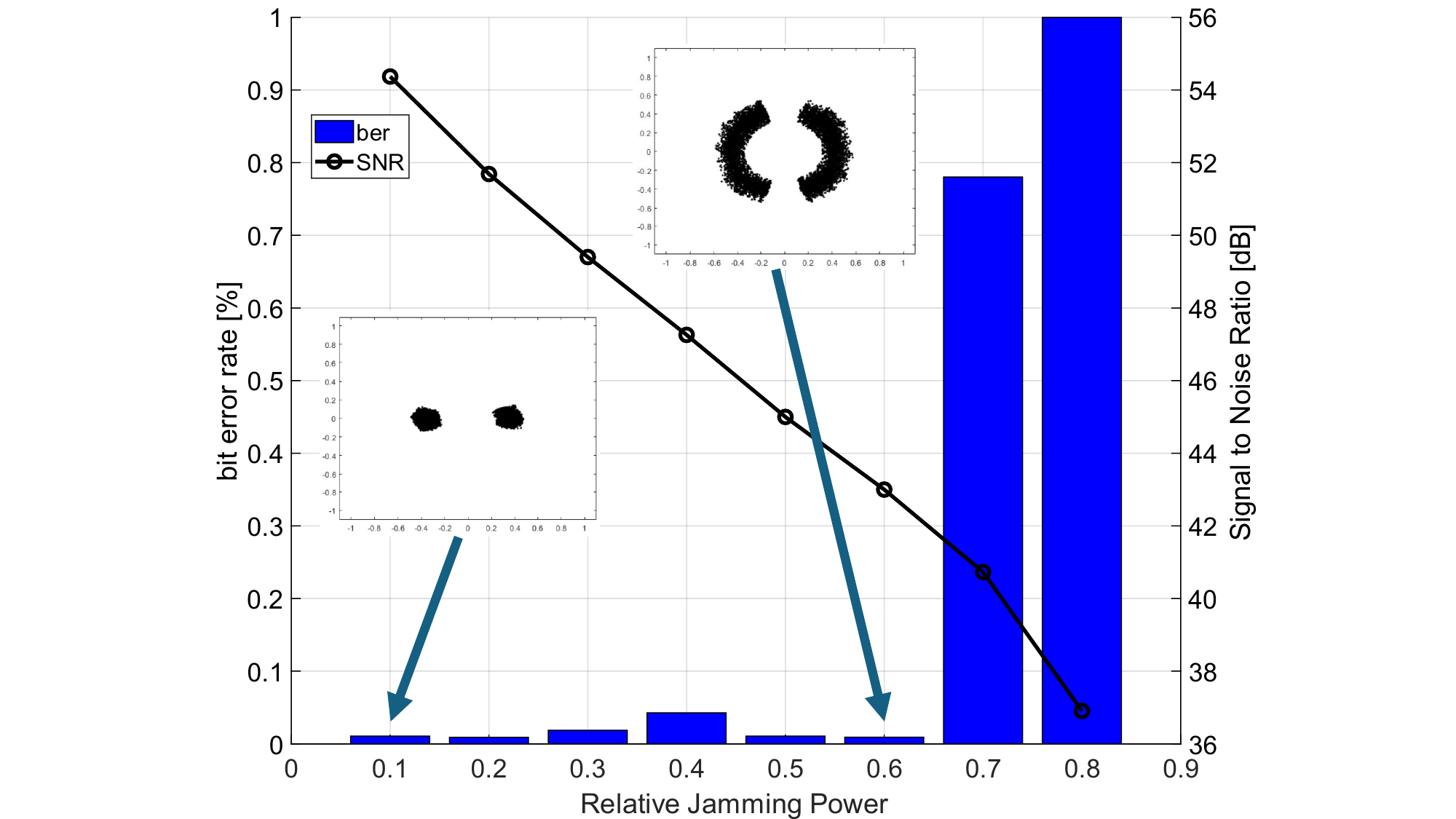}
    \caption{\ac{BER} and \ac{SNR} of a BPSK communication link as a function of the relative jamming power (RJP) injected in the link. The inset figures show the profile of the \ac{IQ} samples at the receiver with RJP values of 0.1 and 0.6. Such a profile can be used to discriminate the presence of the jammer before it affects the quality of the link. RJP refers to the (fraction of the) power respect to the actual power of the transmitter.} 
    \label{fig:ber_snr}
\end{figure}

Several techniques have been designed and tested to detect, identify, and mitigate the presence of a jammer in the radio spectrum. While standard solutions resort to analytics based on \ac{SNR} or other channel state quality metrics, recent results have shown that \ac{DL}-based solutions enable jamming detection even when the jamming signal is low. This is extremely interesting in a scenario where the device moves toward the jammed area: in this context, early jamming detection allows the device to detect the presence of the jammer in advance, when it is still outside the actual jammed area where communication is possible, thus being able to communicate with a remote location, e.g., control center, and taking an informed decision about subsequent movements or actions~\cite{sciancalepore2023_iotj}. 
To prove the viability of channel intelligence, we collected real-world data from a communication link, using as receiver a USRP X310 and as transmitter and jammer two USRP B210 devices, with the jammer deployed farther away from the communication link. The results reported in Fig.~\ref{fig:ber_snr} show that, although the BER of the communication link does not change with the increase of the relative jamming power, the shape of the signal received at our \specint\ receiver does change significantly, enabling early low-BER jamming detection.

Besides jamming detection, channel intelligence applications could also serve the purpose of interference detection and management, e.g., in scenarios where unauthorized devices start transmitting. By monitoring the spectrum, it is possible to detect the presence of anomalous communication patterns and use other solutions, e.g., in the context of Location Intelligence (Sec.~\ref{sec:locInt}) and Communication Intelligence (Sec.~\ref{sec:comInt}), to follow up with the possible threats.

{\bf Eavesdropper intelligence.} The same techniques used to detect the presence of a jammer can also be used to detect passive receiving devices (eavesdroppers). Indeed, as described by Park et al. in~\cite{park2010tcs}, receiving an \ac{RF} signal requires a local oscillator working at the carrier frequency, which leaks a radio signal consistent with its frequency. \specint\ receivers deployed in close proximity could detect the perturbations generated by such devices, thus being able to infer the presence of hidden passive eavesdroppers.

\subsection{Location intelligence} 
\label{sec:locInt}
\specint\ applications falling into the area of \emph{Location Intelligence} allow obtaining insights into the (rough or specific) location of \ac{RF} devices, thus enabling localization, tracking, and inferring their dynamic behavior.

\begin{figure}
    \centering
    \includegraphics[width=.9\columnwidth, angle = 0,trim = 0mm 0mm 0mm 0mm]{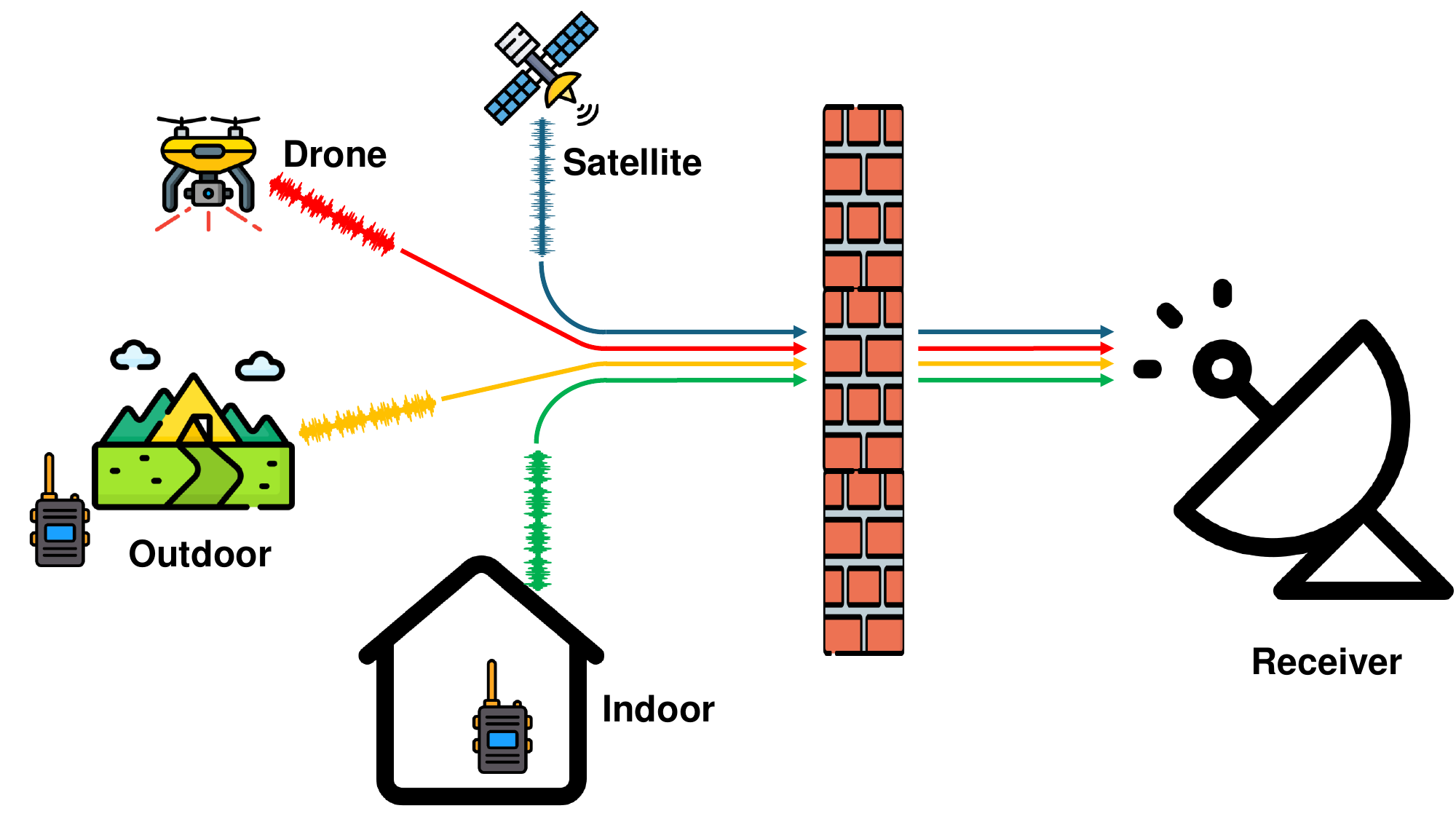}
    \caption{A \specint-enabled receiver can identify the location of transmitters by analyzing the fading process affecting the signals received from such devices.} 
    \label{fig:location_int}
\end{figure}

\emph{Location Intelligence} techniques include research domains such as \ac{RF} anomaly detection, \ac{RF} localization, and \ac{RF} signal forensics. As depicted in Fig.~\ref{fig:location_int}, \ac{RF} signal anomaly detection techniques allow for inferring the rough location of transmitting devices. Indeed, looking at anomalies in the profile of the received signals, it is possible to infer a change in the location of the transmitting device, possibly indicating impersonation and spoofing attacks. One example in this direction is the contribution by Oligeri et al. in~\cite{oligeri2024_sac}, where the authors identify spoofing attacks to satellite systems by detecting anomalies in the fading process affecting the received signal. Since the signal originating from the satellite has different physical-layer characteristics at the receiver side (fading process) compared to the signal originating from a terrestrial source, spoofers emitting satellite \ac{RF} signals could be identified timely and reliably by analyzing the fading process. \emph{Location Intelligence} applications also aid the identification of attacks at the access layer of wireless networks. Indeed, by mapping the regular usage of the \ac{RF} spectrum at various locations, one could detect unusual activity in a specific geographic area covered by one or more \ac{RF} receivers, localizing where the possible intrusion is occurring. By analyzing such anomalies at consecutive time frames, and possibly combining such geospatial information with \emph{Device Intelligence} solutions, it is also possible to track the device generating the anomaly, gaining more insights into the behavior of the attacker and its objectives.

\emph{Location Intelligence} techniques can go even further, by localizing the transmitting \ac{RF} devices to the desired extent. Indeed, by analyzing the \ac{RSS} and \ac{ToA} of various signals by multiple receivers, it is possible to correlate the transmitters and associate events occurring in the network with their presence in a specific area.

Finally, through \emph{Location Intelligence} solutions, it is possible to aid \ac{RF} signal forensics and post-incident analysis. When cyber-physical security breaches occur, e.g., a physical attack on a \ac{CI} or IoT devices, location intelligence can help investigators pinpoint the exact geographical location of the attack source. Also, after the attack, analysts can use \ac{RF} spectrum data with location context to understand how the attack unfolded. For example, if a GPS spoofing attack disrupted the navigation of autonomous vehicles, analyzing the \ac{RF} spectrum around the incident location could provide insight into the methods used by the attackers.

\subsection{Communication intelligence}
\label{sec:comInt}
\specint\ applications falling into the area of \emph{Communication Intelligence} allow obtaining insights into relevant features of a communication channel, e.g., if a signal exists and what is the modulation adopted to communicate the data, representing the first step for successful data decoding.

One of the main research domains falling in this area is \ac{WSI}, comprising all techniques aimed at identifying the transmission parameters of unknown or partially known communication signals~\cite{eldemerdash2016_comst}. \ac{WSI} techniques can process the received signal(s) in space and time to obtain insights into various transmission parameters, e.g., the number of transmitting antennas, their location, orientation, mobility, and the coding scheme adopted by the transmitting source, to name a few. In turn, \ac{WSI} techniques constitute a fundamental building block enabling cognitive radios and networks, that could adapt dynamically to network conditions and environments with minimal coordination and management effort.

A promising subset of \ac{WSI} approaches are the ones for \ac{AMR}, i.e., a set of techniques that allow a receiver to autonomously identify the modulation scheme of an incoming signal, and to configure the related parameters with minimal or no cooperation from the transmitting entity~\cite{zhang2022_dsp}. Although \ac{AMR} has been an active research domain since the early 2000s, it has received renowned attention with the introduction of \ac{DL} techniques, thanks to the remarkable ability of such algorithms to remove noise and extract information from data. Indeed, while recognizing the modulation of a signal can be relatively easy in high \ac{SNR} regimes, it becomes challenging in low-SNR scenarios, where the noise corrupts the signal and makes the (received) physical-layer information (IQ samples) difficult to extract and assess. Transmission systems emitting \ac{RF} signals could also use multiple cooperating antennas, making the overall problem even more challenging to solve. Moreover, the higher the order of the modulation scheme, the harder the identification of the specific modulation scheme, due to the wide range of possibilities available. In this context, \specint\ solutions can use multiple receiving antennas, being mobile, to acquire  specific signals from various locations, and then use \ac{DL} algorithms to identify the modulation scheme that best matches the relevant features of the received signal. In this context, there is already research showing the capability to recover the information about the modulation of a specific signal in challenging scenarios, e.g., with very low \ac{SNR} or when there are limited samples of a specific modulation technique available for training \ac{DL} algorithms. Beyond maximizing performance, \specint\ applications for communication intelligence also have to reduce complexity, and thus, execution time, as much as possible, so to enhance situational awareness capabilities even in complex dynamic \ac{RF} environments, where multiple entities could be transmitting, each with its own modulation scheme. Thus, when coupling \ac{AMR} solutions with powerful parallel computation equipment, monitoring systems (e.g., for Defence) could scale up their surveillance capabilities, enabling fine monitoring and decoding of various simultaneous signals in a target area.

\subsection{Ambient intelligence} 
\label{sec:envInt}
{\em Ambient intelligence} is derived from cyber-spectrum sensing, and it has a wide range of cyber-physical security applications. One significant application is in human activity recognition. By analyzing the \ac{RF} signal patterns, it is possible to identify the presence and movements of humans in close proximity to the transmitter-receiver link. This capability extends to interpreting specific gestures, e.g., stationary or moving people, allowing for non-intrusive monitoring and control systems that respond to human actions without the need for wearable devices~\cite{wu2015_jsac}. Moreover, such applications allow for people movement detection without relying on cameras, thereby maintaining privacy while ensuring that unauthorized access is promptly identified and addressed.

Ambient intelligence applications, in principle, could go even further, by discriminating the type of object crossing an \ac{RF}communication link by analyzing the distortion of the signals at the receiver. Indeed, the distortion caused by humans on \ac{RF} signals at the receiver is different than the distortion caused by objects with a different shape and movement patterns, e.g., drones, possibly enabling detection of more advanced threats.

Similar techniques are conceived in the research domain of \ac{RF} Imaging, which focuses on the use of radio waves to create visual representations of objects and environments. \ac{RF} imaging utilizes electromagnetic waves in the radio frequency spectrum. This technology exploits the ability of radio waves to penetrate various materials, providing unique advantages in scenarios where optical or other imaging techniques cannot work. \ac{RF} imaging has applications across diverse domains, including cyber-physical security, e.g., through-wall surveillance. 

\begin{figure}
    \centering
    \includegraphics[width=.9\columnwidth, angle = 0,trim = 20mm 0mm 30mm 0mm]{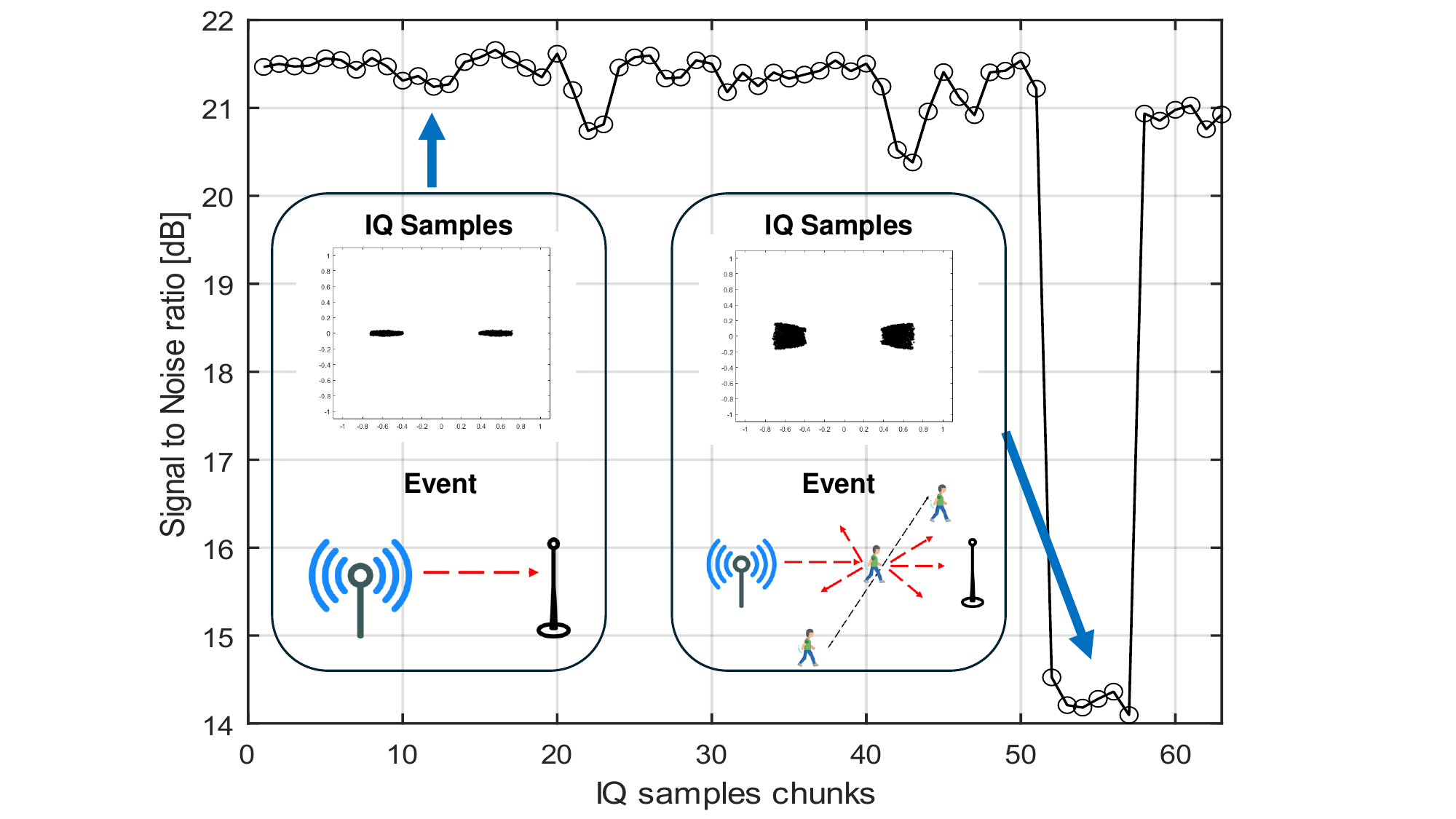}
    \caption{\ac{SNR} and received IQ samples at the receiver when people cross the line-of-sight between the transmitter and the receiver. When people pass through the communication link, IQ samples change their shape, possibly allowing for event detection and classification.} 
    \label{fig:snr_composed}
\end{figure}

Figure~\ref{fig:snr_composed} shows the profile of the RF signal received at our USRP X310 when using the same proof-of-concept deployment described in Sec.~\ref{sec:devInt}. It is worth noting that the drop of the \ac{SNR} at around chunk 50 of our acquisition is correlated with a person crossing the transmitter-receiver line of sight. Looking at the shape of the IQ samples at the receiver, we notice that the clouds are much larger and noisier than usual. We envisage the possibility for future research to investigate how signal distortions can be characterized and leveraged to detect people and objects passing close to the deployment, so developing advanced \emph{Ambient Intelligence} applications.

\section{Cross-Applications Challenges and Road Ahead}
\label{sec:future}

Our preliminary results presented in Sec.~\ref{sec:applications} demonstrate the technical feasibility of \specint\ applications. However, many cross-application challenges still need to be solved to deploy fully reliable \specint\ solutions.

{\bf Noise Mitigation.} The major challenge for \specint\ applications is noise mitigation. Noise can have many forms. For example, background noise is introduced by the hardware of the collection devices and affects the capability of the overall system to detect occurring events. Such noise can be mitigated by choosing more expensive collection devices, characterized by minimal noise figures. However, other noise sources are not that easy to mitigate. For example, in an ambient intelligence application, entities crossing the monitored area can cause sudden fluctuations in channel conditions, preventing the identification of physical events occurring at the same time. Similarly, in a channel intelligence application, a device could start to transmit on the same channel monitored by the system, leading to a false positive jamming detection even when sporadic interferences occur. Thus, we need to design effective and reliable \specint\ applications taking noise into account by design, so to focus on the aspect(s) of interest and reject as many as possible disturbing events causing (small) fluctuations in the channel. 

{\bf Embedded Systems Drifts.} In the last years, many contributions have provided evidence that various side effects affecting embedded systems can cause changes in the way such devices collect information from wireless channels~\cite{alhazbi2023_acsac}. Such changes, if not taken into account at the \specint\ system level, could introduce false positives and false negatives, i.e., detection of events of interest not corresponding to reality and missed detections. Side effects affecting \specint\ applications include, e.g., environmental temperature, aging of the components, fluctuations in the power line, firmware updates, and reboot of the devices. Such events, typical of the regular operation of the collection devices, may lead to a (false) change in the channel conditions perceived by the system. As a result, future research should look into better identifying and understanding such factors. Moreover, we need better mathematical models allowing us to predict the change in the behavior of the collection devices, so as to require minimal intervention by humans and make these systems even more autonomous.

{\bf Training Data Scarcity.} Many \specint\ applications require training data to run reliably. For example, channel intelligence and ambient intelligence applications require data about regular channel conditions to detect anomalies. Such applications could also require data of specific events for detecting their occurrence, e.g., a person crossing the deployment rather than a drone or a bird. However, generating such training data could be challenging in many real-world conditions. Moreover, a challenge might also be represented by simply reproducing the event that we want the \specint\ application to detect, e.g., a bird passing through the deployment. Therefore, future research in this area should focus on enhancing detection capabilities even when the data available for training is minimal, or even absent. In this context, the application of few-shot learning techniques to \specint\ represents an intriguing future development.

{\bf Runtime Resource Overhead.} \specint\ applications require the collection of vast amounts of data and the processing on a powerful computing unit. When the area to monitor is relatively small, the number of data collection devices required to completely cover it can be limited and, as a result, the data can be processed in a limited amount of time, providing almost real-time event detection. However, as the number of collection devices deployed in the field increases, \specint\ applications require exponentially more bandwidth, processing, storage, and possibly energy to meet real-time requirements. To increase the scalability of \specint\ applications, we require even more scalable \ac{DL} algorithms, capable of detecting the patterns of interest from the smallest possible amount of data, exploring the possibly hidden relationships in the data provided by multiple data collection devices.

{\bf New Cross-Domain Applications.} In the previous section, in an attempt to systematize the cyber-physical security application of \specint, we have identified five reference classes. However, such classification might be fuzzy, and cross-domain \specint\ applications may emerge. For example, in a battlefield scenario, troops could use Channel Intelligence to identify the frequencies used by the enemy, jam such frequencies with a custom pattern unpredictable to the enemy, and use the same frequencies (when available) to identify the devices (device intelligence) and detect events (ambient intelligence). Overall, we oversee a range of applications emerging from the use of (or enabled by) \specint, which could revolutionize military operations making them smarter, stealthier, and more effective.

{\bf Individuals' Privacy Preservation.} When applied to civilian deployments, e.g., Critical Infrastructures, \specint\ applications should trade off the capability to monitor the environment with people's privacy. Indeed, \specint\ applications, especially in the context of device and ambient intelligence, could enable the tracking of a user and a possible breach of personal information (for example, visited areas and personal preferences). One way to deal with individual privacy could be to couple \specint\ applications with other systems, e.g., facial recognition, to allow focusing only on suspects while preserving the privacy of trusted ones. 
Alternatively, trusted personnel could feature protective solutions, e.g., the ones proposed in~\cite{irfan2024arxive}, so to avoid tracking.

\section{Conclusions and Future Work}
\label{sec:concl}

In this paper, we introduced the novel paradigm of \specint, i.e., the joint application of next-generation technologies like \aclp{SDR}, \ac{AI}, and parallel processing techniques to detect cyber-physical events of interest from the analysis of the RF wireless spectrum, to provide enhanced real-time cyber-physical security. We identified the research niche occupied by \specint\ approaches at the intersection between various existing research domains, and we identified five main areas of applications for cyber-phsyical security, i.e., Device Intelligence, Channel Intelligence, Location Intelligence, Communication Intelligence and Ambient Intelligence. For some of them, through an actual proof-of-concept using \acp{SDR}, we provided intuition of the feasibility of their application for cyber-physical security. Finally, we identified several research challenges and promising future research directions that could be explored further by Industry and Academia to turn \specint\ research into real-world products deployed to improve cyber-physical security applications. Overall, we believe that our contribution could further stimulate the interest in such technologies and finally unlock the potential of RF spectrum analysis and intelligence for cyber-physical security.


\bibliographystyle{IEEEtran}
\balance
\bibliography{main}

\end{document}